\documentclass[aps,epsfig]{revtex4}
\usepackage{graphicx}
\begin{document}
\def\be{\begin{equation}}
\def\ee{\end{equation}}
\def\bfi{\begin{figure}}
\def\efi{\end{figure}}
\def\bea{\begin{eqnarray}}
\def\eea{\end{eqnarray}}

\title{Fluctuations of two-time quantities and 
time-reparametrization invariance in spin-glasses}
\author{Claudio Chamon}
\affiliation{Department of Physics, Boston University, MA 02215, USA}
\author{Federico Corberi}
\affiliation{Dipartimento di Fisica ``E.R. Caianiello'', Universit\`a di Salerno, 
via Ponte don Melillo, 84084 Fisciano (SA), Italy, and 
Universit\'e Pierre et Marie Curie - Paris 6, 
Laboratoire de Physique Th\'eorique et Hautes Energies, 4 Place Jussieu, 
Tour 13, 5\`eme \'etage, 75252 Paris Cedex 05, France.}
\author{Leticia F. Cugliandolo}
\affiliation{Universit\'e Pierre et Marie Curie - Paris 6, 
Laboratoire de Physique Th\'eorique et Hautes Energies, 4 Place Jussieu, 
Tour 13, 5\`eme \'etage, 75252 Paris Cedex 05, France.
}

\begin{abstract}

This article is a contribution to the understanding of
fluctuations in the out of equilibrium dynamics of glassy systems. By
extending theoretical ideas based on the assumption that
time-reparametrization invariance develops asymptotically we deduce 
the scaling properties of diverse high-order correlation 
functions. We examine these predictions with numerical 
tests in a standard glassy model, the $3d$ Edwards-Anderson spin-glass,
and in a system where time-reparametrization invariance is not expected to hold,
the $2d$ ferromagnetic Ising model, both at low temperatures. Our 
results enlighten a qualitative difference between
the fluctuation properties of the two models and show that scaling properties
conform to the time-reparametrization
invariance scenario in the former but not in the latter.

\end{abstract}

\maketitle


\section{Introduction}

In a series of papers  the idea that time-reparametrization 
invariance (TRI) should be the reason for large non-equilibrium fluctuations in 
glassy systems was introduced.
This claim was substantiated with the analysis of the 
Martin-Siggia-Rose-Janssen-deDominicis (MSRJD) 
action for Langevin stochastic processes~\cite{Chamon02,Castillo02,Charbonneau04,Castillo2} 
and by numerical simulations of various glassy 
models~\cite{Castillo02,Charbonneau04,Jaubert07,Castillo3}. 
The state of the art of these ideas is 
summarized in~\cite{Chamon07}. 

The non-equilibrium situation we are concerned with is a dynamic process  
where a system is  quenched below a transition or dynamic crossover with 
some protocol and it is let relax subsequently. The most interesting time regime 
after the quench is the one in which the size of the sample is the largest 
(though still finite) scale and times are long but not as much as to allow for
equilibration. The system is then in an asymptotic non-equilibrium regime
in which a kind of dynamic scaling is very often realized. In this stage the behavior 
of time-dependent observables is rather well understood~\cite{reviews-mean-field}. 

The sources of fluctuations in the dynamics of these systems are
disorder in the initial conditions, thermal noise, quenched disorder and external
forces, if present.
In~\cite{Chamon07} and references therein the proposal was to
average over any sort of quenched randomness with the purpose of
bringing together disordered systems -- such as spin-glasses,
superconductors, or other -- and conventional glasses -- of atomic,
molecular or other nature. External forces were set to zero and
had no effect on the dynamics.  One is then left with initial
condition and thermally induced fluctuations only.
In order to study their effects, beside the fully averaged value 
$X$ of physical quantities,  
one has to consider the properties of their fluctuating 
parts $\widehat X$, to be defined properly, which may be 
sample dependent. 
The most complete characterization of fluctuations is
then obtained through the probability distribution $P(\widehat X)$.
In particular, since the relevance of fully-averaged two-time quantities 
such as the self-correlation $C(t_1,t_2)$ 
function, the linear autoresponse $\chi(t_1,t_2)$ and their fluctuating parts,
to the issue of TRI was underlined in~\cite{Chamon02,Castillo02,ChaCugYos06},
one is interested in
the probability distributions of $\widehat C$ and $\widehat \chi$ 
studied as a function of $\widehat C$, $t_1$,
$t_2$, the system size and the other parameters in the model.
This is, however, a formidable task.
Less ambitious but still very instructive is to analyze the first moments of the
distributions, such as the generalized variances that we shall define in the body of this paper.
This is, indeed, the route that 
we shall follow in this article. Resorting to a Langevin description, 
in Sec.~\ref{formalism} we set the formalism for continuous variables 
systems and then 
present analytic expressions for 
these variances in terms of the MSRJD fields, deriving 
scaling laws based upon the TRI
scenario. Next, recalling previous 
results on linear and beyond-linear response theory 
\cite{CorLipSarZan10f,LipCorSarZan08,LipCorZan05}, 
we relate these variances to quantities that can be
measured in Ising spin systems in Sec.~\ref{3dea}. In Sec.~\ref{numerical}
we put the TRI predictions to the test in the Edwards-Anderson (EA) model
in $d=3$ by means of extensive numerical simulations.
The results, which refine and extend the analysis 
performed in~\cite{Castillo02}, are consistent with the predictions of TRI.
In the same section we present another set of numerical simulations
performed on a coarsening system, the ferromagnetic Ising model in $d=2$,
for which TRI is not expected to hold~\cite{ChaCugYos06}, showing a different
pattern of fluctuations, already at a qualitative level.
A summary and the conclusions are contained in Sec.~\ref{concl}.  

\section{Formalism for systems with continuous variables and 
time reparametrization invariance} \label{formalism}

\subsection{Generating functional}
\label{sec:generating-functional}  
 
The stochastic dynamics of continuous variables in  a macroscopic system 
are usually described with a Markovian Langevin equation with additive white 
noise. The ensemble of such  coupled stochastic equations can be 
recast into a path-integral called Martin-Siggia-Rose-Jenssen-deDominicis (MSRJD)
generating functional that is better adapted for a general type of  analysis.  This formalism has been 
reviewed in several publications~\cite{MSJ} so we shall not repeat the construction of the 
path-integral here. It reads
\begin{eqnarray}
{\cal Z}[\eta, i\hat \eta]= 
\int {\cal D}\phi {\cal D}i\hat \phi \ e^{-S[\Phi;\eta,\hat \eta]} \ 
P_{IC}[\phi (x,t_0)]
\end{eqnarray}
with 
\begin{equation}
S[\Phi;\eta,\hat \eta] = \int_{t_0}^\infty dt dx
\left\{
i\hat \phi(x,t) [ \dot\phi(x,t) + \frac{\delta V[\phi]}{\delta \phi(x,t)} - h(x,t) ] + T (i\hat \phi(x,t))^2
+\eta(x,t) \phi(x,t) + \hat \eta(x,t) i\hat \phi(x,t) \right\}
\label{eq:action}
\end{equation}
where, to simplify the notation, we focused on a one-dimensional theory described by a scalar
field $\phi $.  $t_0$ is the initial time, when the
system is set in contact with a thermal bath at temperature $T$. We
set $k_B=1$  and we absorb the friction coefficient into a redefinition of time.
$P_{IC}[\phi(x,t_0)]$ is the probability distribution of the field
initial conditions. Henceforth we shall not write the limits of the
time-intervals explicitly.  $V$ is the potential energy from which the
deterministic force in the Langevin equation derives. $h$ is an
external perturbation that couples linearly to the field $\phi$ in the
potential.  $\eta$ and $\hat \eta$ are two space and time dependent
sources.  The path integral runs over all possible configurations of
the field $\phi$ with initial condition $\phi(x,t_0)$. The
measure is defined as ${\cal D}\phi\equiv \prod_{k=0}^{\cal N}
\prod_{a=1}^L d\phi(x_a, t_k)$ and we jump over subtle discretization
issues that are discussed in many publications by recalling that we do
not need to worry about any determinant contribution to the action.
The integral over the auxiliary field $i\hat\phi(x,t)$ is similarly
discretized, namely ${\cal D}i\hat\phi\equiv \prod_{k=1}^{\cal N}
\prod_{a=1}^L di\hat\phi(x_a, t_k)$.  Without loss of generality for our 
purpose, we 
take $P_{IC}$ to be flat. We use a compact
notation for the two fields, $\Phi=(\phi \;,\; i\hat \phi)$, in the action and hereafter we 
label $\Phi^0=\phi$ and $\Phi^1=i\hat\phi$ its components.  Due to the normalization
of the thermal noise distribution the path-integral is identical to
one, ${\cal Z}[0,0]=1$, in the absence of sources;
this feature will be useful later.  Noise averages translate into averages 
computed within the path-integral formulation with weight $e^{-S[\Phi;0,0]}$. 

For definiteness, and to parallel the discussion in~\cite{Chamon07} and references therein (to which we refer for further details),
let us separate a 
quadratic term from the non-linear one, $V_{nl}$, in the potential 
\begin{equation}
V[\phi] = \frac{1}{2} \int dx dy \ J(x,y) \phi(x,t) \phi(y,t) + V_{nl}[\phi]
\end{equation}
and assume that the symmetric time-independent exchange matrix $J$ is 
made of independent elements each of them
Gaussian-distributed $P[J] \propto \prod_{xy} \exp[-(1/2) \int dx dy \ J^2(x,y) K^{-1}(x,y)]$
with 
$K^{-1}$ the connectivity matrix. For a lattice model with first nearest-neighbours 
and no self-interactions, as the $3d$ Edwards-Anderson spin-glass that we 
studied numerically, $K^{-1}(x,y)=J^2/z$ for nearest-neighbours and zero otherwise, 
with $J$ the exchange constant and $z$ the connectivity of the lattice.
Quenched disorder is treated 
on average by exploiting the fact that ${\cal Z}[0,0]=1$. Concretely, one averages the
dynamic generating functional over the $J$-distribution with no need to use the replica
trick. This gives rise to 
a quartic term in the action of the form
\begin{eqnarray}
S_4[\Phi]&=&-\frac{1}{8} \int dx dy \int dt_1 dt_2 \ 
\left[
i\hat\phi(x,t_1)\,\phi(y,t_1) + \phi(x,t_1)\,i\hat\phi(y,t_1)
\right]
K(x,y) 
\left[
i\hat\phi(x,t_2)\,\phi(y,t_2) +\phi(x,t_2)\,i\hat\phi(y,t_2) 
\right]
\nonumber\\
&=&-\frac{1}{8} \int dx dy \int dt_1 dt_2 \ 
\Phi^a(x,t_1)\Phi^{\bar a}(y,t_1)\, K(x,y)\, \Phi^b(x,t_2)\Phi^{\bar b}(y,t_2) 
,
\end{eqnarray}
where $a,b=0,1$ and $\overline 0=1$ and $\overline 1=0$ (here repeated
indices are summed). This term can be decoupled by introducing four spatially local two-time fields $Q^{ab}(x;t_1,t_2) $,
$a,b=0,1$, 
with a Hubbard-Stratonovich transformation. One then has
\begin{eqnarray}
S_4[\Phi] &= &
\int {\cal D}Q 
\exp \left[ 
\frac{1}{2} \int dx dy \ K^{-1}(x,y) \int dt_1 dt_2 \ Q^{ab}(x;t_1,t_2)  Q^{\overline b\overline a}(y,t_2,t_1) 
\right.
\nonumber\\
&&
\qquad\qquad\qquad\qquad \left.
-
\int dx \int dt_1 dt_2 \ \Phi^a(x,t_1) Q^{ab}(x;t_1,t_2) \Phi^b(x,t_2) 
\right]
\;.
\label{eq:actionQ}
\end{eqnarray}
The zero-source generating functional can now be written as
${\cal Z} [0,0] = \int {\cal D}Q {\cal D}\Phi \ e^{-S[Q,\Phi]}$
where the action $S[Q,\Phi]$ includes all terms, those in Eq.~(\ref{eq:actionQ})  that
depend on $Q$ and those in 
Eq.~(\ref{eq:action}) with $V$ replaced by $V_{nl}$ that depend on $\Phi$ only.

\subsection{Relation between composite operators of $Q$ and $\Phi$ fields}
\label{sec:transformations}  

The TRI arguments \cite{Chamon07}, which are
briefly recalled in Sec.~\ref{sec:TRI}, rely on the transformation properties of the
fluctuating $Q$ fields that, in turn, translate into relations between the
average values of products of the $Q$ fields. These objects, though,
are auxiliary in the calculation and are not
directly accessible in a simulation or experiment. 
In order to proceed towards a test of TRI the next step, to which
this paragraph is devoted, is to relate the statistical
properties of the Hubbard-Stratonovich fields $Q$ to those of the
field $\Phi$. 
In particular, our program is to find 
relations between averages of products of $Q$s and moments of the $\Phi$s
from which  directly measurable correlation
and response functions will be constructed.  To this end 
we perform an infinitesimal translation of the field $Q\to Q+\delta Q$, 
which induces a change in the action $S[Q,\Phi]\to S[Q,\Phi]+\delta S[Q,\Phi]$,
with
\begin{eqnarray}
\delta S [Q,\Phi] &=& - \int dx \int dt_1 dt_2 \ \Phi^a(x,t_1) \Phi^b(x,t_2) \delta Q^{ab}(x;t_1,t_2) +
\int dx dy \ K^{-1}(x,y) \int dt_1 dt_2 \ \delta Q^{ab}(x;t_1,t_2)  Q^{\overline b\overline a}(y,t_2,t_1) 
\nonumber\\
&& \qquad \qquad +
\frac{1}{2} \int dx dy \
K^{-1}(x,y) \int dt_1 dt_2 \ \delta Q^{ab}(x;t_1,t_2)  \delta Q^{\overline b\overline a}(y,t_2,t_1)
\; . 
\end{eqnarray}
The normalization ${\cal Z}[0,0]=1$ implies $\langle e^{-\delta S} \rangle=1$ 
where $\langle \cdots \rangle = \int {\cal D}\Phi {\cal D}Q \
\cdots \; e^{-S[\Phi,Q]}$. 
To first and second order in $\delta Q$ one has 
\begin{equation}
\int dy \ K^{-1}(x,y) \ \langle Q^{\overline b\overline a}(y;t_2,t_1) \rangle= 
\langle \Phi^a(x,t_1) \Phi^b(x,t_2) \rangle
\; , 
\label{eq:average-dQ}
\end{equation}
and
\begin{eqnarray}
- K^{-1}(x,u) \  \delta_{a\overline d}  \ \delta_{b \overline c} \ \delta(t_1-t_4) \ \delta(t_2-t_3)  
&=&
\left\langle \left( \Phi^a(x,t_1) \Phi^b(x,t_2) - \int dy \ K^{-1}(x,y) \ Q^{\overline b\overline a}(y;t_2,t_1) \right)
\right.
\nonumber\\
&& 
\qquad \times \left.
\left(
\Phi^c(u,t_3) \Phi^d(u,t_4) - \int dv \ K^{-1}(u,v) \  Q^{\overline d\overline c}(v;t_4,t_3) \right)
\right\rangle \; ,
\label{eq:average-dQdQ}
\end{eqnarray}
respectively. The first equation tells us that the average of the
Hubbard-Stratonovich $Q$ field, coarse-grained over the region in
which $K^{-1}(x,y)$ is different from zero, is equal to the
correlations of two $\Phi$ fields. These in turn are related to 
the usual correlation and linear response functions, as will be
recalled in Sec.~\ref{sec:observables}.
Note that the rhs of Eq.~)\ref{eq:average-dQ}) is symmetric 
under the exchange of $(a,t_1)$ and $(b,t_2)$ and this implies 
that the average in the lhs should also be.
Since we are not only interested in the average of $Q$, but also in their generalized
variances (which are given by averages of four fields, as
discussed in Sec.~\ref{sec:observables}), the content of the second order 
equation~(\ref{eq:average-dQdQ}) is also relevant.
Indeed, the expansion of the product in the
right-hand-side of the second equation yields three kinds of terms, one of which 
is precisely the average of the product of four $\Phi$ fields. Another one is the average
of the product of two $Q$s. The last cross term  is the average of the product of two 
$\Phi$ and one $Q$ fields. Using standard  manipulations of the path-integral,
this term can also be recast into averages over products of $Q$ fields:
\begin{equation}
\langle \Phi^a(x,t_1) \Phi^b(x,t_2)  Q^{cd}(v;t_3,t_4) \rangle =
\langle Q^{cd}(v;t_3,t_4) Q^{\overline b \overline a}(x;t_1,t_2) \rangle -
\delta(t_1-t_3) \delta(t_2-t_4) \delta_{ac} \delta_{bd}  \delta(x-v)
\; .
\label{eq:cross-average}
\end{equation}
Putting everything together,  Eq.~(\ref{eq:average-dQdQ})
allows one to write the average of four $\Phi$ fields as sums of terms with 
averages of products of two spatially coarse grained $Q$ fields. 

As it will be discussed in Sec.~\ref{sec:TRI}, TRI provides
predictions on the transformation properties of composite $Q$-fields
operators under a certain class of transformations. Because we have
just shown above that averages of $Q$ operators are equal to moments
of $\Phi$ fields, we arrive at the following important relation: {\it all
physical observables which can be written in terms of moments of $\Phi$ fields
should inherit the symmetry properties of the $Q$ fields.}

\subsection{Time-reparametrization}
\label{sec:TRI}

An approximate treatment of the MSRJD action~\cite{Chamon02,Castillo02,Castillo2}, 
inspired by the 
solution to mean-field glassy models and mode-coupling 
approximations~\cite{reviews-mean-field}, 
is based on the assumption that it can be separated into two kinds of 
terms, involving fast and slow components of the fields. In the slow contribution
time-derivatives are neglected, integrals involving products of fast and slow contributions
are approximated, and the action is fully written in terms of the 
$Q$ fields only.  The $a$, $b$ indices are linked in such a way that the ensemble of slow 
terms are invariant under the global transformation $t\to h(t)$,
where $h(t)$ is any positive and monotonic function of time.  
This transformation is a time-reparametrization. In terms of the $Q$ fields it implies 
\begin{equation}
Q^{ab}(x;t,s) 
\to Q_{TR}^{ab}(x;t,s) =\left( \frac{dh(t)}{dt} \right)^{a} \left( \frac{dh(s)}{ds} \right)^{b} Q^{ab}(x;t,s)
\; ,
\label{eq:transformation-TRI}
\end{equation}
and, for the averages of $Q$-field products  
\begin{eqnarray}
&& 
\langle Q^{a_1b_1}(x_1;t_1,s_1) \dots Q^{a_nb_n}(x_n;t_n,s_n)\rangle 
\to \langle Q_{TR}^{a_1b_1}(x_1;t_1,s_1) \dots Q_{TR}^{a_nb_n}(x_n;t_n,s_n)\rangle=
\nonumber\\
&& 
\qquad
\left(\frac{dh(t_1)}{dt_1}\right)^{a_1} \left(\frac{dh(s_1)}{ds_1}\right)^{b_1}
\dots 
\left(\frac{dh(t_n)}{dt_n}\right)^{a_n} \left(\frac{dh(s_n)}{ds_n}\right)^{b_n}
\langle 
Q^{a_1b_1}(x_1;t_1,s_1) \dots Q^{a_nb_n}(x_n;t_n,s_n)
\rangle \; .
\label{eq:transf-prod-Q} 
\end{eqnarray}
Equation~(\ref{eq:transf-prod-Q}) describes how a generic
composite $Q$-field operator transforms under time-reparametrization. 
In general, quantities characterized by the same set of indexes $a_1,b_1,\dots,a_n,b_n$
transform in the same way under the time-reparametrization change.
A particular case is that of quantities that are left unchanged
by the transformation, that is to say, TRI-quantities.
These can be obtained by integrating a composite operator such as the ones
entering Eq.~(\ref{eq:transf-prod-Q}) over each time associated to an
$a_k=1$ or $b_k=1$ subscript. Indeed, the prefactors are eliminated by a change of integration variable 
$t \to h(t)$. For the cases $n=1$ and $n=2$,      
recalling Eqs.~(\ref{eq:average-dQ}) and (\ref{eq:cross-average}),
one has that averages of products of $2n$ $\Phi$-fields are TRI
after integration over the times pertaining to the $\Phi ^0$ fields.
The autocorrelation
function $C$ and the associated linear susceptibility $\chi$, as well as
the second moments of their fluctuating parts, belong to this class,
as it will be further discussed in Sec.~\ref{sec:ave-and-fluc}. 
These quantities are the subject of the numerical study carried out in Sec.~\ref{3dea}.  

\subsection{Variances}
\label{sec:observables}

In this Subsection we define the quantities we study in this paper, both analytically 
and numerically. 

\subsubsection{Fully averaged quantities and their fluctuating parts}
\label{sec:ave-and-fluc}

The connected 
two-point two-time correlation function is 
\begin{equation}
C(\vec x,\vec y;t,s) = 
\langle\; (\phi(\vec x,t) -\langle \phi(\vec x,t)\rangle) \;
(\phi(\vec y,s) -\langle \phi(\vec y,s)\rangle)\; \rangle  
\end{equation}
where $\langle \dots \rangle = \int {\cal D}\phi {\cal D}i\hat \phi \dots e^{-S}$.
The averaged linear response of any observable is defined as its variation with 
respect to an applied perturbation, in the limit in which the latter vanishes.
The most studied linear response is the one of the averaged field itself
with respect to a perturbation $h$ which couples linearly to $\phi $ in the Hamiltonian
(a magnetic field, regarding $\phi $ as a spin). A simple calculation shows that,
within the MSRJD formalism, it is simply related to an average of the 
product of the fields $\phi$ and $i\hat \phi$:
\begin{equation}
R^{(1,1)}(\vec x,\vec y;t,t_1) 
= 
T \left. \frac{\delta \langle \phi(\vec x,t)\rangle_h}{\delta h(\vec y,t_1)}\right|_{h=0} 
= 
T \langle \phi(\vec x,t) i\hat\phi(\vec y,t_1) \rangle 
\; . 
\end{equation}
Note that a factor $T$ has been added in the definition of the linear response to ease
the notation here and in the following.
The integral of this quantity is the dynamic susceptibility
\begin{equation}
\chi^{(1,1)}(\vec x,\vec y;t,s) = \int_s^t dt_1 \ R^{(1,1)}(\vec x,\vec y;t,t_1)
\; , 
\end{equation}
which in the following will be more simply denoted as $\chi$, omitting the superscript
$^{(1,1)}$.
We define a particular second order response, $R^{(2,2)}$, as the variation of the 
averaged composite field $\phi(\vec x,t) \phi(\vec y,t)$ with respect to two instantaneous 
perturbations applied at $t_1$ and $t_2$ on the same spatial points 
on which the fields are evaluated. Again, a simple calculation allows one to show 
that within the MSRJD formalism  this quantity equals an average of, in this case,
four fields in the absence of the perturbation:
\be
R^{(2,2)}(\vec x,\vec y;t,t_1,t_2)
\equiv 
T^2 
\left .
\frac{\delta^2\langle\phi (\vec x,t)\phi(\vec y,t)\rangle_h}
{\delta h(\vec x,t_1)\delta h(\vec y,t_2)}\right|_{h=0}
=
T^2 \langle 
\phi(\vec x,t) i\hat\phi(\vec x,t_1) 
\phi(\vec y,t) i\hat\phi(\vec y,t_2) 
\rangle 
\; .
\label{r2var}
\ee 
In general we define generalized responses $R^{(n,m)}$ as the 
response of a composite operator
containing the product of $n$-fields with respect to $m$ magnetic fields.
These are functions of $n+m$ spatial positions and $n+m$ times,
but to keep the notation simple we will omit writing coinciding arguments (for instance we will write
$R^{(1,1)}(\vec x;t,t_1)$ instead of $R^{(1,1)}(\vec x,\vec x;t,t,t_1)$ and
$R^{(2,2)}(\vec x,\vec y;t,t_1,t_2)$ instead of $R^{(2,2)}(\vec x,\vec y,\vec x,\vec y;t,t,t_1,t_2)$).
Quite generally, a variation with respect to the field $h$ can be replaced
by a response field $i\hat \phi$ to compute higher-order response functions and, 
according to the convention introduced above, we include $m$ factors of 
$T$ in the definition of the generalized response functions. 

The next step is to introduce a couple of local fluctuating quantities,
$\widehat C(\vec x,\vec y;t,s)$ and $\widehat \chi(\vec x,\vec y;t,s)$, with the property 
$\langle \widehat C(\vec x,\vec y;t,s)\rangle=C(\vec x,\vec y;t,s)$ and 
$\langle \widehat \chi(\vec x,\vec y;t,s)\rangle=\chi (\vec x,\vec y;t,s)$, to be 
interpreted as the {\it fluctuating parts} of $C$ and $\chi$. Since different
fluctuating quantities can share the same average, this can be done in 
different ways. A detailed discussion of this issue is exposed
in \cite{CorLipSarZan10f}. For continuous variables described by 
a Langevin equation, recalling the formalism of  
Sec.~\ref{formalism}, one is naturally led to consider 
\begin{eqnarray}
&& \widehat C(\vec x,\vec y;t,s)
=
\phi(\vec x,t)\phi(\vec y,s)
\; ,\nonumber
\\
&& \widehat \chi(\vec x,\vec y;t,s)
=
T \int_s^t dt_1 \ \phi(\vec x,t)i\hat \phi(\vec y,t_1)
\; , 
\label{eq:fluc_parts}
\end{eqnarray}
as being the variables whose MSRJD average yield $C$ and $\chi$. 
Still, $\widehat \chi$ involves
the auxiliary field $i\hat \phi$ and, in consequence, it is not directly
accessible in an experiment or a simulation. This problem can be bypassed
by studying the probability distribution of $(\widehat C,\widehat \chi)$
through its moments, since, as we show below, these
can be expressed in terms of the field $\phi$ alone relating them to 
generalized response functions.
Starting from the lowest non-trivial moments, with the definitions (\ref{eq:fluc_parts})
one can build the {\it variances}  
\begin{eqnarray}
V^{CC}(\vec x,\vec y;t,s)& = &\langle \widehat {\delta C}(\vec x;t,s)\widehat {\delta
   C}(\vec y;t,s)\rangle \; ,
\label{eq:VCC}
\\
V^{C\chi}(\vec x,\vec y;t,s)&= &
\langle \widehat {\delta
   C}(\vec x;t,s) \widehat {\delta \chi}(\vec y;t,s)\rangle \; ,
\label{eq:VCchi}
\\
 V^{\chi\chi}(\vec x,\vec y;t,s)&= &\langle \widehat {\delta \chi}(\vec x;t,s) \widehat
 {\delta \chi}(\vec y;t,s)\rangle \; ,
\label{eq:Vchichi}
\end{eqnarray}
where for a generic quantity $\widehat
 X$ we have defined the fluctuation $\widehat {\delta X}=\widehat X- X=
\widehat X-\langle \widehat X\rangle$. 
As anticipated above, the variances containing $\widehat {\delta \chi}$
can be expressed in terms of the 
{\it physical} field $\phi$ alone, by expressing them in terms of 
generalized responses.
For instance, the variance (\ref{eq:Vchichi}) can 
be related to a second-order response, as given by
 Eq.~(\ref{r2var}), 
 \begin{eqnarray}
 V^{\chi\chi}(\vec x,\vec y;t,s) &=& 
 \int_s^t dt_1  
\int_s^t dt_2 \
[ R^{(2,2)}(\vec x,\vec y;t,t_1,t_2) - R^{(1,1)}(\vec x;t,t_1)R^{(1,1)}(\vec y;t,t_2) ]
\; .
 \end{eqnarray}
Analogously, the co-variance $V^{C\chi}$ can be expressed as 
 \begin{eqnarray}
 V^{C\chi}(\vec x,\vec y;t,s) &=& 
T  \int_s^t dt_1 \
 \langle \;
 [\phi(\vec x,t) \phi(\vec x,s)-\langle\phi(\vec x,t) \phi(\vec x,s)\rangle]\;
[\phi(\vec y,t) i\hat\phi(\vec y,t_1)-\langle\phi(\vec y,t) i\hat\phi(\vec y,t_1)\rangle]
\;\rangle
\nonumber\\
&=& 
 \int_s^t dt_1 \
 [
 R^{(3,1)}(\vec x,\vec y;t,s,t_1) 
 -
 C(\vec x;t,s) 
 R^{(1,1)}(\vec y;t,t_1) 
 ],
 \end{eqnarray}
where $C(\vec x;t,s)=C(\vec x,\vec x;t,s)$ is the autocorrelation function and
$R^{(3,1)}(\vec x,\vec y;t,s,t_1)=
\langle \phi(\vec x,t) \phi(\vec x,s)\phi(\vec y,t) i\hat\phi(\vec y,t_1)\rangle$. 
In this paper, we enforce these relations to compute the
moments~(\ref{eq:VCC})-(\ref{eq:Vchichi}) numerically. Higher order moments
are exceptionally demanding to compute, and we leave their study to
further works.
In order to improve the statistics of the
concrete numerical measurements it will be convenient to compute the
double spatial integral:
\begin{equation}
V^{XY}_{k=0}(t,s) \equiv L^{-d} \int d^dx \int d^dy \ V^{XY}(\vec x,\vec y;t,s) \; ,
\label{eq:k=0}
\end{equation}
where $V^{XY}$ is a compact notation to denote the variances $V^{CC}$,
$V^{C\chi}$ and $V^{\chi\chi}$, simultaneously, $d$ is the spatial
dimension and $L^{d}$ is the volume of the system.
A comment is now in order. In spite of the name we gave to it,
$V^{\chi\chi}$ is not the variance of a physical quantity; indeed,
being linked to  response functions, it can even take negative values. Still, in the following
we shall broadly use the term {\it co-variances} (or simply variances)
when referring to $V^{CC}$, $V^{C\chi}$ and $V^{\chi\chi}$.

Let us now come back to what TRI implies for the quantities introduced above,
following the prescriptions of Sec.~\ref{sec:TRI}.
We start from the usual correlation $C(\vec x,\vec y;t,s)$ and linear response $\chi (\vec x,\vec y;t,s)$.
The former is written in terms of two $\phi$ fields, in terms of $Q$'s it is then 
a function of $Q^{00}$ only and it is invariant under 
Eq.~(\ref{eq:transformation-TRI}).   
$\chi $
involves, besides one field $\phi$, also one $i\hat\phi$ field, but
there is an integral over the time argument of the latter; thus $\chi$
does not transform either. 
Proceeding analogously for the variances,
$V^{CC}$ involves only
$\phi$ fields, $V^{C\chi}$ and $V^{\chi\chi}$
involve one and two fields $i\hat\phi$ respectively,
together with one and two integrations over the corresponding
time variables. Then, also the variances are TRI,
similarly to $C$ and $\chi$.

\subsubsection{An equilibrium  sum rule}

A sum rule that generalizes FDT to higher  powers is easily proven 
using the invariance of the action and the measure under the transformation 
$\phi(\vec x, t) \to \phi(\vec x, -t)$ and 
$i\hat\phi(\vec x, t) \to i\hat\phi(\vec x, -t)+\beta \partial_t \phi(\vec x, -t)$
in equilibrium~\cite{Arbicu}. Indeed, 
\begin{equation}
0 = 
\left\langle
\left[ \int d^d x \  \widehat {\delta \chi}(\vec x; -t,-s) \right]^n
\right\rangle
=
\left\langle 
\left[ \int d^d x \left(  \widehat {\delta \chi}(\vec x; t,s) 
- \widehat{\delta C}(\vec x; t,t) +  \widehat{\delta C}(\vec x; t,s)
\right) \right]^n
\right\rangle
,
\end{equation}
for $t\geq s$.
The first identity is due to causality and the second one is derived 
by using the symmetry. 
As a special case one has 
\begin{equation}
V_{k=0}^{CC} + 2 V_{k=0}^{C\chi} + V^{\chi\chi}_{k=0}=0, 
\label{eq:sumrule}
\end{equation} 
valid for $n=2$ and for variables such that $\widehat{\delta C}(\vec x; t,t)=0$,
e.g. Ising spins.
 
\subsection{Restricted averages}
\label{subsubsec:restricted}

For a generic fluctuating observable $\widehat X$, besides the
standard full average $X=\langle \widehat X \rangle$ we also introduce
the restricted average $X_{\widehat C}=\langle \widehat X\rangle_{\widehat C}$,
namely an average taken only over the instances (realizations) with a
given value $\widehat C$ of the fluctuating autocorrelation, as
suggested in \cite{CorCug09f}.
Clearly one has 
\be X=\sum _{\widehat C}X_{\widehat C} \ P(\widehat C), 
\ee
where $P(\widehat C)$ is the marginal probability that a chosen
dynamical trajectory has autocorrelation $\widehat C$.  For a
co-variance between two quantities $\widehat A$ and $\widehat B$ one
has 
\be V^{AB}=\sum _{\widehat C}V^{AB}_{\widehat C} P(\widehat C)+{\cal V}^{A B}
\label{compos_var}
\ee where $V^{AB}_{\widehat C}=\langle \widehat A \widehat
B\rangle_{\widehat C}-\langle \widehat A\rangle_{\widehat C} \langle
\widehat B\rangle_{\widehat C}$ is the restricted covariance of
$\widehat A$ and $\widehat B$ (for $\widehat C$ fixed), and ${\cal V}^{A
  B}=\sum _{\widehat C} A_{\widehat C} B_{\widehat C} P(\widehat C)-
\sum _{\widehat C} A_{\widehat C} P(\widehat C) \  \sum_{\widehat
  C}B_{\widehat C} P(\widehat C)$ is the covariance of the restricted
averages $A_{\widehat C}$ and $B_{\widehat C}$ as $\widehat C$ is varied.

\subsection{Scaling}
\label{sec:scaling}

In the late stage of the evolution of slowly relaxing systems,
some sort of dynamical scaling is usually obeyed. Indeed this is what we will
find in our simulations. In this Section, we discuss the implications 
of TRI in such systems.
Let us start considering a couple of generic local observables $A(t,s)$, $B(t,s)$, 
such as the autocorrelation $C$ or the autoresponse $\chi$, with scaling expressions
\begin{eqnarray}
A(t,s)&=&\xi(s)^{b_A}f_A\left [\frac{\xi (t)}{\xi (s)}\right ],\\
B(t,s)&=&\xi(s)^{b_B}f_B\left [\frac{\xi (t)}{\xi (s)}\right ],
\end{eqnarray}
where $\xi (t)$ is a certain function of time. We emphasize that we always
refer to the slow degrees of the dynamical process and that any explicit
fast contribution must be subtracted out if present.
In some cases, as in coarsening systems,
$\xi $ can be readily interpreted as a growing correlation length, although this is
not crucial in what follows.
Clearly, if the two quantities transform in the same way under TRI it must be $b_A=b_B$
and $f_A(z) \propto f_B(z)$ in the double limit $s\to\infty$ and 
$z\to\infty$, i.e. $t\gg s$.

If one time, say $t$, can be eliminated in favor of $B$, the quantity $A$ can also be 
expressed as $A(t,s)=\widetilde A(B,s)=\xi (s)^{b_A}f_{AB}[\xi(s)^{-b_B}B]$, 
where $f_{AB}=f_Af_B^{-1}$, and one can introduce the
{\it slope}
\be
X_{AB}(B,s)=\frac{\partial \widetilde A(B,s)}{\partial B} =
\left . \xi(s)^{-b_B+b_A} \ \frac{d f_{AB}(z)}{d z}\right \vert _{z=\xi(s)^{-b_B}B} .
\label{eq:X}
\ee
Under time-reparametrization $t\to h(t)$
one has
\begin{eqnarray}
A(t,s)&\to& A_{TR}(t,s)=\xi(h(s))^{b_A} \ f_A\left[\frac{\xi (h(t))}{\xi (h(s))}\right]
, \\
B(t,s)&\to& B_{TR}(t,s)=\xi(h(s))^{b_B} \ f_B\left[\frac{\xi (h(t))}{\xi (h(s))}\right] \label{eq:B}.
\end{eqnarray}
Defining $\widetilde A_{TR}$ similarly to what we did for 
$\widetilde A$, one has  
$\widetilde A_{TR}(B_{TR},s)=\xi(h(s))^{b_A}f_{AB}\left [\xi(h(s))^{-b_B}B_{TR}\right ]$,
and
\be
X_{A_{TR}B_{TR}}(B_{TR},s)=\frac{\partial \widetilde A_{TR}(B_{TR},s)}{\partial B_{TR}}=
\left . \xi(h(s))^{-b_B+b_A} \ \frac{df_{AB}(z)}{dz}\right \vert _{z=\xi(h(s))^{-b_B}B_{TR}}.
\label{eq:X_TR}
\ee
Comparing Eqs.~(\ref{eq:X}) and (\ref{eq:X_TR}) we conclude that,
if the two quantities transform in the same way, namely
$b_A=b_B$, the slope of the original  parametric curve, $\chi_{AB}(B,s)$,  
evaluated at a value $B$ is equal to the slope of the transformed parametric
curve, $\chi_{\widetilde A_{TR}B_{TR}}(B_{TR},s)$,  
evaluated  at $B_{TR}=[\xi(s)/\xi(h(s))]^{-b_B}B$.
In other words, under time-reparametrization 
$B$ and $A$ are transformed in such a way as to preserve the original
curve with slope $X_{AB}$, although different parts of the curve are 
related by the transformation.
Considering the self-correlation and the self-response, 
$A=\chi$ and $B=C$, $X_{\chi C}$ turns out to be the so-called fluctuation-dissipation
ratio $X=T/T_{eff}$, $T_{eff}$ being usually denoted as the {\it effective temperature}
(here we simply quote the term, without entering into the delicate problem of its 
interpretation as a physical temperature)~\cite{Teff}. In this case, since according to the predictions
of TRI $\chi$ and $C$ are both invariant under time-reparametrization, 
Eq.~(\ref{eq:X_TR}) implies a finite $T_{eff}$. On the other hand, in systems where
TRI is not obeyed $C$ and $\chi$ may transform differently and this prevents $T_{eff}$
from being finite. This leads to the conjecture that TRI
may not be obeyed in coarsening systems with an infinite $T_{eff}$, as it 
was analytically shown in a solvable case in \cite{ChaCugYos06}. 
For the Ising model this issue will
be discussed in Sec.~\ref{comparison}.

Let us consider now non-local quantities such as, for instance
$V^{XY}(\vec x,\vec y;t,s)$. 
Since spatial translational invariance as
well as isotropy are respected in the mean, they are actually
functions of $r=|\vec x-\vec y|$ only. 
Assuming scaling, one has
\be
V^{XY}(\vec x,\vec y;t,s)=\xi(s)^{b_{XY}} \ f_{V^{XY}}\left[\frac{r}{\xi(s)},\frac{\xi(t)}{\xi(s)}\right] .
\label{eq:scalD}
\ee
The $k=0$ component of such a quantity is then expected to scale as
\begin{equation}
V^{XY}_{k=0}(t,s) \equiv L^{-d} \int d^dx \int d^dy \ V^{XY}(\vec x,\vec y;t,s)=
\xi (s)^{d+b_{XY}} \ f_{V^{XY}_{k=0}}\left [\frac{\xi(t)}{\xi(s)}\right ].
\label{eq:scalDk}
\end{equation}
In Sec.~\ref{numerical} we shall study  the behavior of 
the variances $V^{XY}_{k=0}$, hereafter simply referred to as $V^{XY}$, numerically. 
Since TRI predicts that all the $V^{XY}$ should  transform in the same way, we expect 
them to scale as
\begin{equation}
V^{XY}(t,s) = \xi(s)^{b_{XY}} \ f_{XY}\left [\frac {\xi(t)}{\xi(s)}\right ],
\label{eq:scalVk}
\end{equation}
with a {\it unique} exponent $b_{XY}=b$.

\section{Spin systems} 
\label{3dea}

In this Section we adapt the definitions of 
the variances and restricted averages presented in the previous section to 
discrete variables systems. 

\subsection{Basic definitions} \label{secbasic}

\subsubsection{Models}

We consider a spin system described by 
the Hamiltonian
\be
H(\sigma )=-\sum _{\langle ij\rangle}J_{ij}\sigma _i \sigma _j,
\label{ham}
\ee
where $\sigma _i=\pm 1$ are spin variables 
located on the $N=L^d$ sites $i$ of a cubic lattice of linear size $L$.
$\sigma $ denotes a spin configuration and $J_{ij}$
are the couplings between nearest neighbor sites $\langle ij\rangle$.
A constant $J_{ij}=J$ defines the ferromagnetic Ising model while 
in the EA model we choose random bimodal couplings, $J_{ij}=\pm J$, 
with zero mean. In the following we shall assume $J=1$. 
The evolution of the system is governed by the transition rates
$w(\sigma '\vert \sigma)$ 
for going from a configuration $\sigma $ to another $\sigma '$.
In the following we consider the dynamics whereby single spins are updated,
namely $w(\sigma '\vert \sigma )=(1/N)\sum _i w_i({\sigma '\vert \sigma })$.
$w_i$ is the transition rate for flipping $\sigma _i$, and
$\sigma$ and $\sigma '$ may differ in the $i$-th spin only.
We use $w_i$ of the Glauber form $w_i=(1/2)[1+\tanh (-\Delta E/2T)]$, where 
$\Delta E$ is the energy variation due to the proposed spin flip and $T$ is 
the temperature of the heat bath.

\subsubsection{Averaged two-time functions}

The autocorrelation function is defined as \be C(t,s)=\langle \sigma
_i(t)\sigma _i(s)\rangle.  \ee Here we have implicitly assumed
$\langle \sigma _i(t)\rangle \equiv 0$ $\forall t$, and $\langle \dots
\rangle = \overline {\langle \dots \rangle _J }$ denotes, 
for the spin glass model, 
the double average over thermal histories and initial conditions (denoted by
$\langle \dots \rangle _J$) and over the quenched realization of the
random couplings $J_{ij}$ (denoted by $\overline
{\cdots} $); for the ferromagnetic Ising model only the first average is present.  
The associated susceptibility is 
\be 
\chi(t,s)=\int _s^t
dt_1\ R^{(1,1)}(t,t_1),
\ee 
where 
\be R ^{(1,1)}(t,s)
=
T \left .\frac{\delta \langle \sigma
  _i(t)\rangle} {\delta h_i(s)}\right \vert _{h=0} 
\ee 
is the
impulsive auto-response function and $h_i(s)$ is an instantaneous
magnetic field switched on and off at time $s\le t$. These averaged
quantities do not depend on the position $i$ and we thus omit the $i$
subscript.

\subsubsection{Fluctuating two-time quantities}

In order to study fluctuating two-time quantities in spin models and 
to put the TRI to the test, we have
to provide expressions for the variances (\ref{eq:VCC})-(\ref{eq:Vchichi}) in systems with 
discrete variables, where the path integral representation discussed in
Sec.~\ref{sec:generating-functional} is not available. Following \cite{CorLipSarZan10f},
this can be done by using an extension of the fluctuation-dissipation theorem
\cite{LipCorZan05} which allows one to relate response functions of any order
to correlation functions in systems out of equilibrium.
To linear order one has     
\begin{equation}
\widehat \chi_i(t,s)=
\frac{1}{2}\left[\sigma _i(t)\sigma_i(t)- \sigma _i(t)\sigma_i(s)- 
\sigma_i(t)  \int _{s}^t dt_1 B_i(t_1)\right ],
\label{1.5}
\end{equation}
where $B_i=-\sum _{\sigma '}[\sigma _i-\sigma' _i] 
w_i(\sigma '\vert \sigma)$. It easy to show that this operator rules the 
evolution of the local magnetizations, namely 
$d\langle \sigma _i(t)\rangle/dt=\langle B_i(t)\rangle$. 
It can also be shown \cite{LipCorZan05} that the expression (\ref{1.5})
holds also in a Langevin description if the analogue of $B_i$, 
namely the deterministic force $-\delta V/\delta (\phi(x,t))$, is used.
This strengthens the relation between the discrete variable formalism
of this section and the one introduced in Sec.~\ref{formalism}.
With this fluctuating susceptibility and $\widehat C_i(t,s)=\sigma _i(t)\sigma _i(s)$
one readily builds $V^{CC}$ and $V^{C\chi}$. 
The procedure to determine the expression of $V^{\chi\chi}$ is 
more subtle as we now discuss.
First, this quantity is related to the
second order response in Eq.~(\ref{r2var}), which for discrete spins 
on a lattice is more properly
written as 
\be
R^{(2,2)}_{ij}(t,t_1,t_2)\equiv 
T^2 \left .
\frac{\delta^2\langle\sigma_i(t)\sigma_j(t)\rangle_h}
{\delta h_{i}(t_1)\delta h_j(t_2)}\right|_{h=0},
\label{app1.2}
\ee
by
\be
V^{\chi\chi}_{ij}(t,s)\equiv \int_{s}^t dt_1\int_{s}^t dt_2 \left [R^{(2,2)}_{ij}(t,t_1,t_2)
-R^{(1,1)}(t,t_1)R^{(1,1)}(t,t_2) \right ].
\label{chi2r2}
\ee
Proceeding as before, we express $R^{(2,2)}$ 
in terms of correlation functions through the 
extension of the
fluctuation-dissipation theorem mentioned above.
Using the expression for the second order response~(\ref{chi2r2}) 
derived in~\cite{LipCorSarZan08} one arrives at~\cite{CorLipSarZan10f}
\begin{eqnarray}
V^{\chi\chi}_{ij}(t,s)&=&\frac{1}{4}
\langle\sigma_i(t)\sigma_j(t)[\sigma_i(t)-\sigma_i(s)][\sigma_j(t)-\sigma_j(s)]\rangle
- \frac{1}{4} \int_{s}^t dt_2 \ \langle\sigma_i(t)\sigma_j(t)[\sigma_i(t)-\sigma_i(s)]B_j(t_2)\rangle
\nonumber \\
&&-\frac{1}{4} \int_{s}^t dt_1 \ \langle\sigma_i(t)\sigma_j(t)B_i(t_1)[\sigma_j(t)-\sigma_j(s)]\rangle
+ \frac{1}{4} \int_{s}^t dt_1 \int_{s}^t dt_2 \ \langle\sigma_i(t)\sigma_j(t)B_i(t_1)B_j(t_2)\rangle
\nonumber \\
&& +\frac{1}{4}\delta_{ij}\int_{s}^t dt_1 \ 
\langle\sigma_i(t)^2 \tilde{B}_i(t_1)\rangle
-\chi(t,s)\chi(t,s),
\label{app1.4}
\end{eqnarray}
where $\tilde{B}_i=-\sum_{\sigma'}[\sigma_i'-\sigma_i]^2w(\sigma'|\sigma)$.
In this way all the variances are written in terms of the spin configurations
and they can be computed in a numerical experiment.
This method to compute response functions 
in the unperturbed evolution [i.e. Eqs.~(\ref{1.5}) and (\ref{app1.4})], 
besides the advantage of having
the vanishing perturbation limit built in, is by far more efficient \cite{algo} as
compared to the standard method where the perturbation is actually
switched on, particularly for higher order responses, as discussed 
in \cite{CorLipSarZan10}.
The accuracy of the results presented in Sec.~\ref{numerical} would not have
been possible without resorting to this technique.
In the following we focus on the variances' $k=0$ component, 
defined analogously to Eq.~(\ref{eq:k=0}),
\be
V^{XY}(t,s)=\frac{1}{N} \sum _{ij}V^{XY}_{ij}(t,s)
\; , 
\label{kappazero}
\ee
where to ease the notation we omitted the under-script $k=0$.  
 Notice that $V^{CC}$ is also (apart
from an overall factor $N$) the variance of the space average of the
autocorrelations, $\widehat C_r=(1/N)\sum _i \widehat C_i$, and it is
hence bound to be positive.  This quantity is related to the one
measured in~\cite{Castillo02} with the only difference that here we
take the coarse-graining linear size, $\ell$, to be equal to the size
of the system, $\ell=L$.  We recall that in equilibrium the variances
are related by Eq.~(\ref{eq:sumrule})~\cite{CorLipSarZan10f}.

\subsubsection{Restricted averages}

The definitions in sec.~\ref{subsubsec:restricted} can be straightforwardly applied to problems with 
discrete variables. 

\section{Numerical simulations} \label{numerical}

In this section we present the results obtained  
with Monte Carlo simulations of the $3d$ Edwards-Anderson model and the
$2d$ ferromagnetic Ising model. We choose to present, first, the bare data 
for the $3d$ EA model and their scaling analysis. Only later we interpret these
forms in terms of TRI. While doing so we confront to the $2d$ Ising model data.

\subsection{$3d$ Edwards-Anderson spin glass} \label{EAnumeric}

\subsubsection{Equilibrium behavior} \label{seceq}

In this section we study the $V^{XY}$s when the spin glass is in
equilibrium at $T=0.8$. The knowledge of their behavior will be useful
in the interpretation of the non-equilibrium results, described in
Sec.~\ref{secneq}. With this perspective in mind, we selected very small sample
sizes to allow for a rapid equilibration. We simultaneously monitored
finite-size effects as explained below. These conditions led us to
work with $L\leq 5$. In the following, time is
measured in Monte Carlo steps (MCs).

\begin{figure}[h]
    \centering
   \rotatebox{0}{\resizebox{.7\textwidth}{!}{\includegraphics{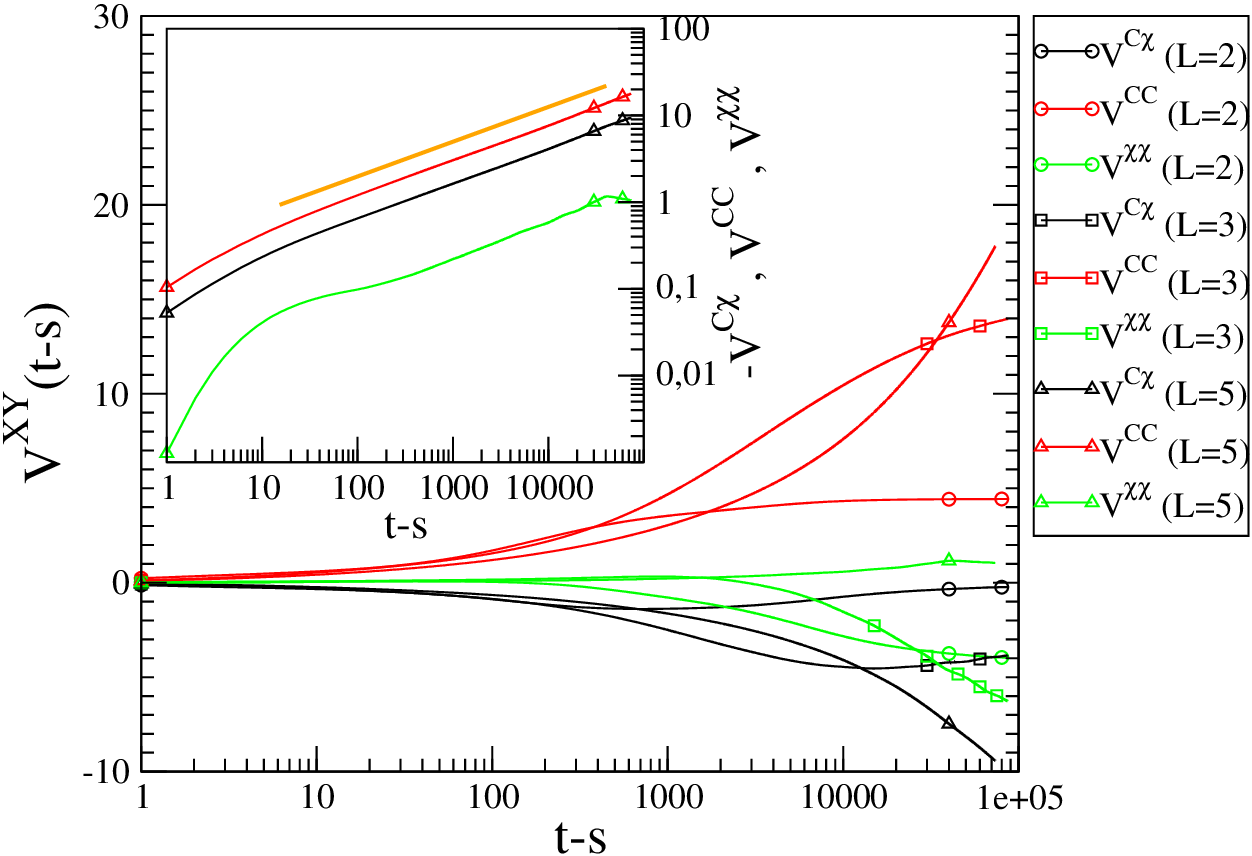}}}
   \vspace{0.5cm}
   \caption{(Color online.) Monte Carlo simulations of the $3d$ EA model. 
   In the main part of the figure the zero wave-vector variances $V^{XY}$ are 
plotted against $t-s$ in a log-linear scale,
for systems of linear sizes $L=2,3,5$ in equilibrium at $T=0.8$. See the key for the color and 
symbol code.
Averages are taken over $2.2$x$10^8$ realizations of the random interactions and 
the initial conditions. In the inset the data for
$L=5$ are shown in a log-log scale
(we use $-V^{C\chi}$ in order to display a positive quantity). The bold orange
straight segment represents 
the power law $(t-s)^\alpha$ with $\alpha=0.4$.}
\label{fig_eq_1}
\end{figure}

The qualitative behavior of the $V^{XY}$s and the role of the system
size can be understood by inspection of Fig.~\ref{fig_eq_1}.  
Equilibrium dynamics is stationary so the three
$V^{XY}$ depend on the time-difference only. 
$V^{CC}$
is positive, for the reasons discussed in Sec.~\ref{secbasic}.
Initially (soon after a microscopic time) it grows as a power-law,
$V^{CC}\sim (t-s)^\alpha$, and after a characteristic time difference
$t-s\sim t_M(L)$ it flattens and attains a constant value
$V^{CC}_\infty(L)=(1/N)\sum _{ij}\overline{(\langle \sigma _i\sigma
  _j\rangle_J)^2}$.  This shows that the $L$-dependence of
$V^{CC}_\infty(L)$ is a finite size effect due to $L<\xi _{eq}$.
Because of the rapid increase of the characteristic time $t_M$ with
$L$, the asymptotic value is fully reached only for $L=2$ in our
simulations.  For $L=5$ the flattening is completely out of reach.
From the analysis of this dynamic curve we extract the value $\alpha=0.4$ for
the power-law exponent.

\vspace{1.5cm}

The covariance $V^{C\chi}$ is negative. This can be understood since
the equilibrium parametric plot $\chi (C)$ is a curve with local negative slope. 
Apart from the sign, this quantity behaves initially as $V^{CC}$, exhibiting
a power-law growth with an exponent that is compatible with the one
found for $V^{CC}$. Around $t_M$ the quantity $V^{C\chi}$ reaches a minimum
and then grows towards an asymptotic value 
$V^{C\chi}_\infty$ which can be shown to be 
$V^{C\chi}_\infty=0$ in cases in which the system is not magnetized, 
$\langle \sigma _i\rangle=0$  \cite{CorLipSarZan10f}. 
Indeed, in the case $L=2$, where
$t_M$ is sufficiently short to let the system reach the asymptotic value, one has a very small
final numerical value of $V^{C\chi}_\infty$.

Finally,  $V^{\chi\chi}$ is initially positive and grows proportionally to the other two
variances. Around $t_M$ it reaches a maximum and then decreases to a negative
asymptotic value given by $V^{\chi\chi} _\infty=-V^{CC}_\infty$~\cite{CorLipSarZan10f}.  

In short, the initial time-difference dependence is given by 
\begin{equation}
V^{XY}(t,s) = V^{XY}(t-s) \simeq (t-s)^\alpha \;\;\;\; \mbox{for} \;\;\;\; t-s \ll t_M(L)
\; . 
\end{equation}

Let us stress that curves with different $L$ collapse in this power-law regime.
Hence, for instance, the curves obtained for $L=5$ are also representative
of a system with $L>5$ at least up to times $t\simeq 10^5$ MCs.
This fact will be used in the next section, where a sample with $L=10$
will be studied up to $t=10^4$ MCs. The sum rule (\ref{eq:sumrule}) is satisfied at
all times within our numerical accuracy.

\vspace{1.5cm}
\begin{figure}[h]
    \centering
   \rotatebox{0}{\resizebox{.7\textwidth}{!}{\includegraphics{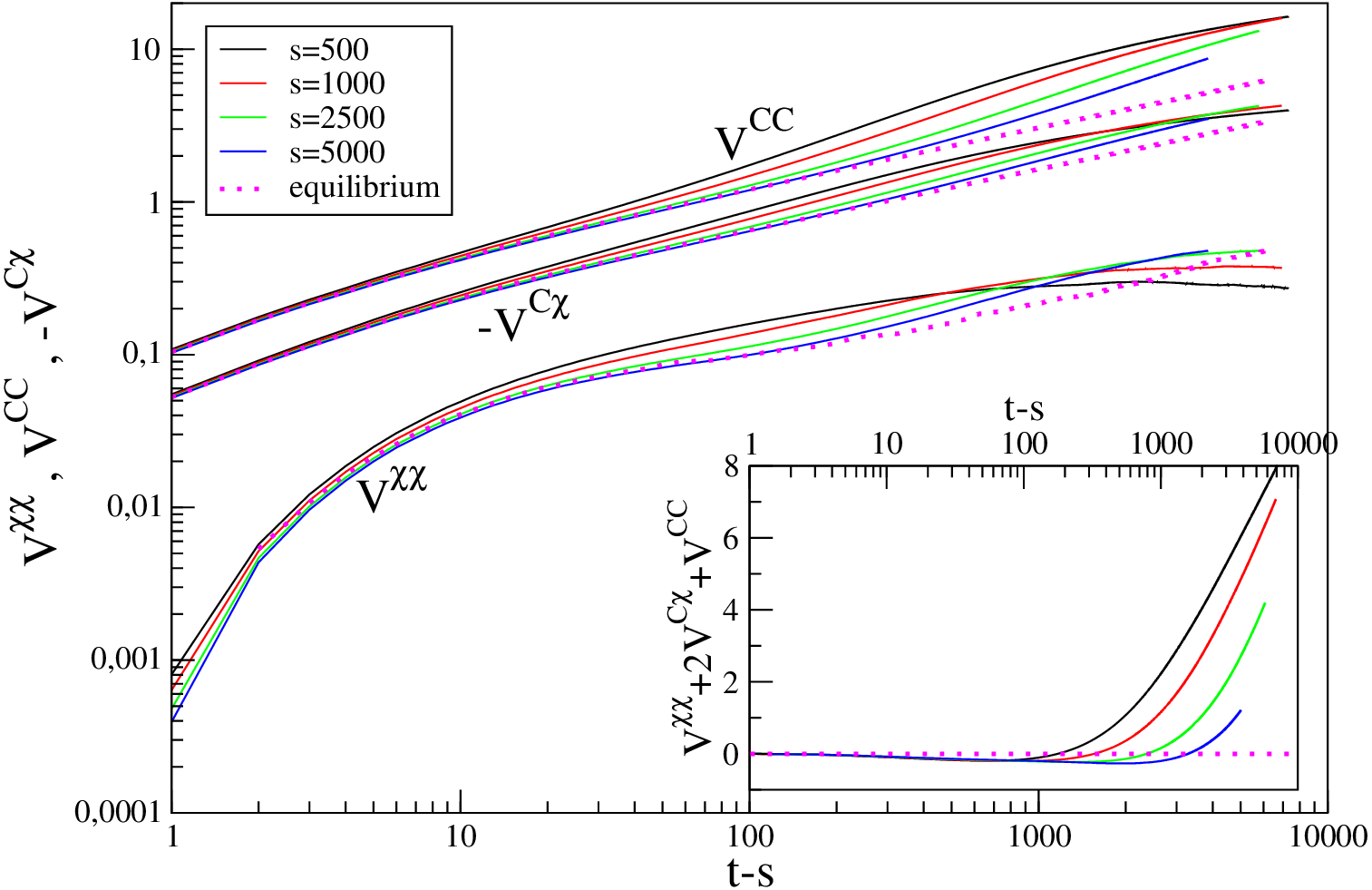}}}
   \vspace{0.5cm}
   \caption{(Color online.) The zero wave-vector variances $V^{XY}$ 
after a quench of the $3d$ EA model 
from $T\to\infty$ to $T=0.8$. In the main panel the data are plotted against 
$t-s$ in a log-log scale for different values of the waiting-time $s$ given in the key.
The three groups of curves are $V^{\chi\chi}$, $-V^{C\chi}$ 
and $V^{CC}$ from
bottom to top, respectively.
Averages are 
taken over $6.3$x$10^7$ realizations of the randomness and initial conditions. 
The (purple) dotted curves are the
equilibrium behavior in the case with $L=5$. In the inset the quantity 
$V^{\chi\chi}+2V^{C\chi}+V^{CC}$ is plotted against $t-s$, proving that it vanishes 
at short time differences.  Thus the equilibrium relation applies in the quasi stationary 
regime.} 
\label{fig_neq_0}
\end{figure}

\subsubsection{Out of equilibrium behavior: variances} \label{secneq}

We now turn to the study of $V^{XY}$ in out of equilibrium conditions.
We performed a quench from $T\to\infty$ to the working
temperature $T=0.8$ at time $t=0$ and we followed the relaxation
of a system of size $L=10$ up to $10^4$ MCs.  We checked
that in the range of times accessed in our simulations the system is
free from finite size effects and remains out of equilibrium. This is
probably due to the fact that, although some kind of growing length
$\xi$ is present in the
system~\cite{others-length,Jaubert07,janus}, in the range of times
considered here it is still of the order of a few lattice spacings
(say 3 or 4 at most).  The choice of the system size and range of
times is dictated by the necessity to tame the colossal noise
afflicting $V^{\chi\chi}$ in order to have sufficiently clean
data. Indeed, increasing the system size or running for longer time enhances the
noise. Concerning size, this is due to the fact that only the
couples $ij$ within a distance of order $\xi$ in
Eq.~(\ref{kappazero}) contribute significantly to the sum,
as expressed by Eq.~(\ref{eq:scalD}), while the
others introduce noisy contributions (in spite of having zero
average).  Regarding time, as $\xi$ increases the number
$(L/\xi)^3$ of independent regions decreases and this also
enhances the noise in the course of time.  
Furthermore, being two-time correlations, the $V^{XY}$ decay to zero
as $t-s$ increases, making the signal/noise ratio worse.
The data presented below
are obtained by averaging over more than $6.3$x$10^7$ realizations of
thermal histories, initial conditions and coupling constants.

The behavior of the variances is shown in Fig.~\ref{fig_neq_0}.
One finds $V^{CC}>0$ and $V^{C\chi}<0$, basically for the same reasons
discussed in Sec.~\ref{seceq}, while $V^{\chi\chi}$ is positive. 
For small time differences
$t-s\lesssim \tau(s) \sim s$ one finds a quasi-equilibrium behavior: 
the variances depend on the time difference $t-s$ only,
they superimpose on the curves found in equilibrium 
(for the largest size $L=5$, see the discussion at the end of Sec.~\ref{seceq})
and the relation (\ref{eq:sumrule}) is verified (see the inset). In short, one has 
\be
|V^{XY}(t,s)|\simeq (t-s)^\alpha \;\;\;\;\; \mbox{with} \;\;\;\;\;
\alpha \simeq 0.4 \;\;\;\;\;
\mbox{in the stationary regime.}
\label{eq-scaling}
\ee

\begin{figure}[h]
    \centering
   \rotatebox{0}{\resizebox{.7\textwidth}{!}{\includegraphics{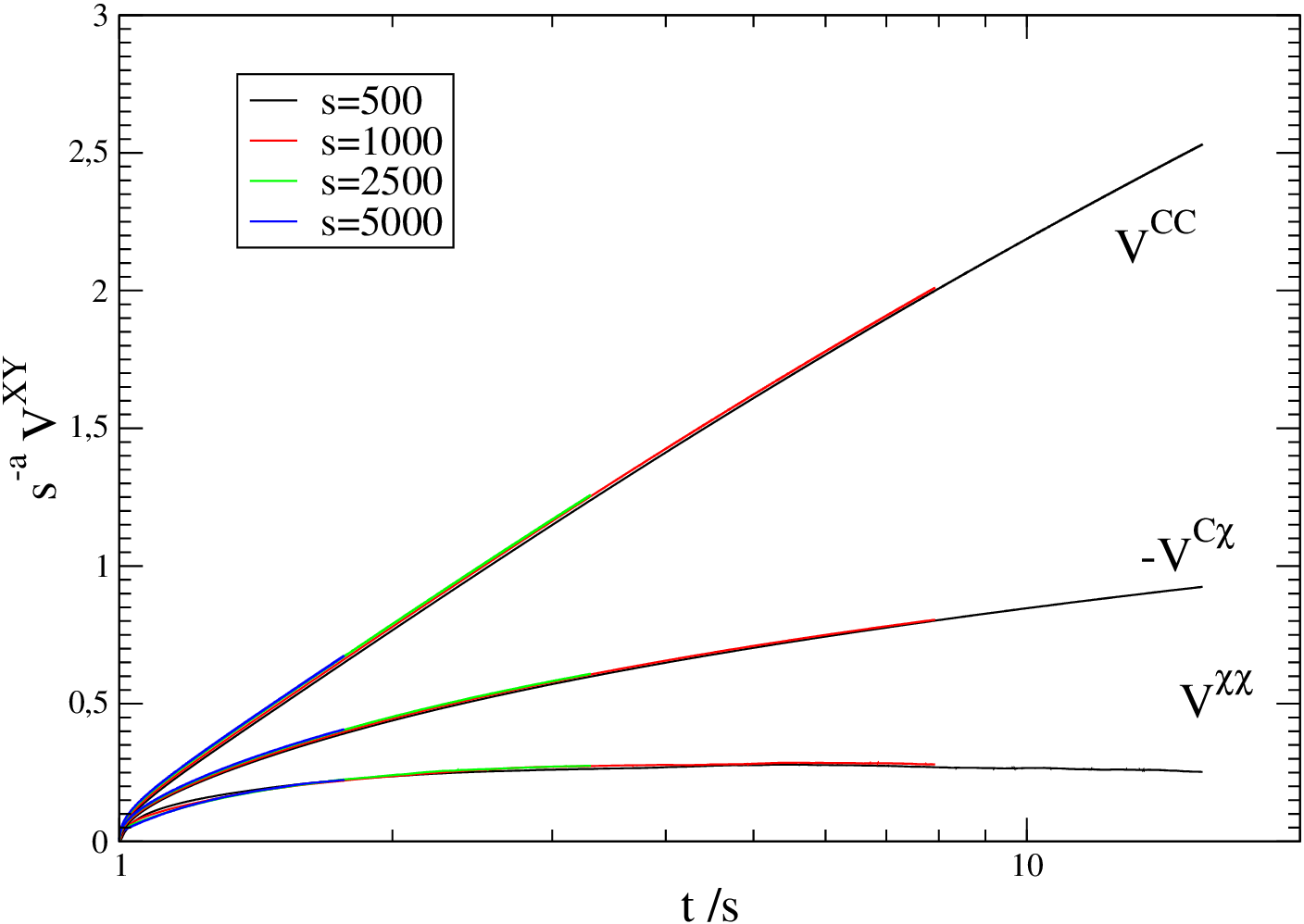}}}
     \vspace{0.5cm}
   \caption{(Color online.)  The bare data shown in Fig.~\ref{fig_neq_0} are 
collapsed by plotting $s^{-a}V^{XY}$ (with $a =0.3$) against 
$t/s$ in a log-linear scale for different values of $s$ (see the key).
The three groups of curves are $6V^{\chi\chi}$, $-1.5V^{C\chi}$  and $V^{CC}$ from
bottom to top, respectively.  The factors -1.5 and 6 are used to achieve a better 
presentation of the data since $V^{CC}>|V^{C\chi}|\gg V^{\chi\chi}$. 
}
\label{fig_neq_1}
\end{figure}

For larger values of $t-s$ the system enters the aging regime in which
non-equilibrium effects become important. One observes an explicit
dependence on the two times, and a departure from the equilibrium
behavior. In this regime, the $V^{XY}$ start growing faster than the
power-law characteristic of the equilibrium behavior. For still larger
values of $t-s$, $V^{CC}$ and $V^{C\chi}$ continue their growth
steadily but in a slower logarithmic way (see Fig. \ref{fig_neq_1}),
$\simeq (\ln t)^{b_{XY}}$. The exponents $b _{XY} $ were found to be
$b_{CC}= 0.86$ and $b_{C\chi}= 0.52$ in the region $\ln (t/s)>1$. On
the other hand, $V^{\chi\chi}$ seem to saturate to a constant value
(or there could be a maximum after which it would decay, we cannot
exclude this possibility). The difference of the scaling function with
the ones of the other variables remains, however, only logarithmic
and, in some sense, marginal. Notice that the
equilibrium part of the variances is not negligible in the aging
regime. Instead, increasing as a power-law, it is expected to be
asymptotically dominant (although we cannot access such a long time
behavior in the simulations), since $V^{XY}$ either converge to a
constant value ($V^{\chi\chi}$) or grow logarithmically ($V^{CC}$ and
$V^{C\chi}$).

\vspace{1.5cm}
\vspace{0.5cm}

Interestingly,  as  shown in 
Fig.~\ref{fig_neq_1}, all variances obey the scaling form
\be
V^{XY}(t,s) \simeq s^{a} \ f_{XY}(t/s)
\;\;\;\;\;
\mbox{in the aging regime,}
\label{scaling}
\ee
with the {\it same} exponent $a \simeq 0.3$, the same dependence on the two times $t$ and 
$s$ through their ratio in the remaining factor, but different scaling functions.
This fact has important consequences towards the tests of 
the TRI scenario as we shall discuss in Sect.~\ref{numvsrep}.

\vspace{1.5cm}
\begin{figure}[h]
    \centering
   \rotatebox{0}{\resizebox{.7\textwidth}{!}{\includegraphics{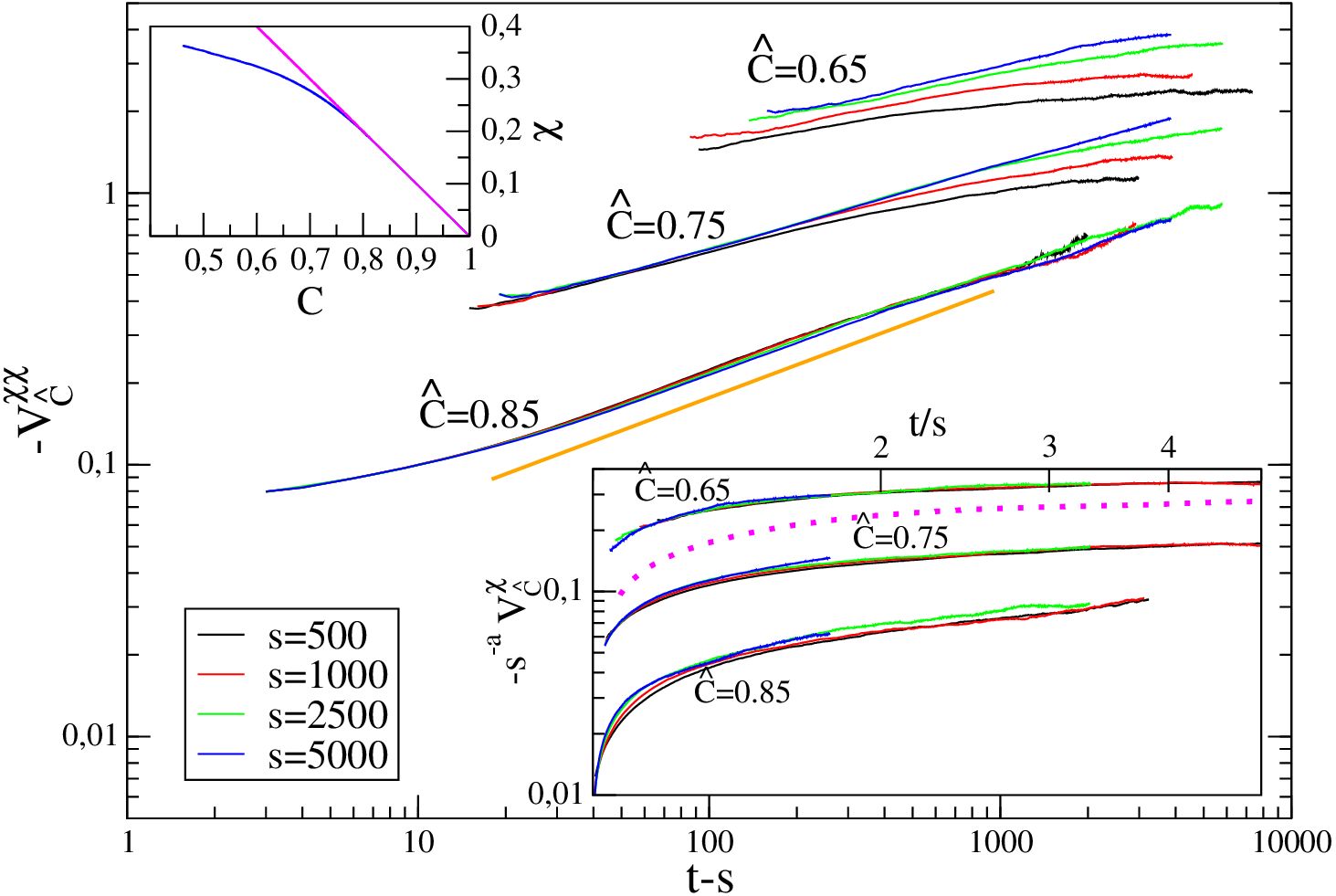}}}
   \vspace{0.5cm}
   \caption{(Color online.) 
Restricted averages $-V^{\chi\chi}_{\widehat C}$  in the $3d$ EA model 
for different values of
$\widehat C$ indicated in the figure 
and $s$ given in the key. The data are shown in a log-log scale. 
The three groups 
of curves have been vertically displaced for convenience.
The (orange) straight  segment is the power law $(t-s)^\alpha$, with $\alpha =0.4$.
In the upper-left inset the parametric
plot of $T\chi$ versus $C$ is plotted for $s=5\times 10^3$ MCs. The straight (magenta) line
is the equilibrium behavior $T\chi = 1-C$.
In the lower-right inset the same curves of the main part of the figure
are plotted with the rescaling of Fig.~\ref{fig_neq_1}, namely by
plotting $s^{-a}V^{\chi\chi}_{\widehat C}$, with $a=0.3$, against $t/s$ (in a log-log scale). The 
dotted 
(magenta) line is $s^{-a}V^{\chi\chi}$, for $s=5\times 10^2$ MCs.} 
\label{fig_neq_3}
\end{figure}

\subsubsection{Out of equilibrium behavior: restricted averages} \label{secneq2}

The restricted average of the susceptibility, $V^{\chi\chi}_{\widehat
  C}$, is shown in Fig.~\ref{fig_neq_3} for three values of the
two-field product, $\widehat C=0.85$, $\widehat C=0.75$, and $\widehat
C=0.65$.  Interestingly enough, for all $\widehat C$ the sign of the
restricted variance is negative, namely the opposite of the sign of the
whole $V^{\chi\chi}$.  From the parametric plot of $\chi $ versus $C$
(included in the upper inset) one concludes that the value $\widehat
C=0.85$ belongs to the quasi-equilibrium regime, where $T\chi \simeq
1-C$, whereas $\widehat C=0.65$ is far into the aging regime. The
value $\widehat C=0.75$ is somehow in between. For the choice
$\widehat C=0.85$ the curves for different values of $s$ collapse,
showing that $V^{\chi\chi}_{\widehat C}$ depend only on $t-s$ and,
moreover, the data are well described by the power-law
$V^{\chi\chi}_{\widehat C}\sim (t-s)^\alpha$ with an exponent that is
consistent with $\alpha =0.4$ as found for the full variance
$V^{\chi\chi}$ in equilibrium and in the quasi-stationary regime. 
Therefore for values of $\widehat C$ in
the quasi-equilibrium regime of the parametric plot $\chi$ vs. $C$,
the restricted averages have an equilibrium behavior (even for times
$t-s>s$ where the whole $V^{\chi \chi}$ falls out of
equilibrium). Moving towards the aging regime, for $\widehat C=0.75$
and $\widehat C=0.65$ the data collapse is progressively lost,
starting from the largest values of $(t-s)/s$.  Interestingly, while
time translation invariance is lost, the scaling of the data
transform into the non-equilibrium one,
Eq.~(\ref{scaling}), found for the totally averaged variances, with
the same exponent $a\simeq 0.3$. This is shown in the lower inset of
Fig.~\ref{fig_neq_3}, where we scaled the restricted averages by
plotting $s^{-a}V^{\chi\chi}_{\widehat C}$ against $t/s$. The collapse is quite
good for $\widehat C=0.65$ and it gets worse for increasing $\widehat
C$. The reason why the collapse for $\widehat C=0.85$, although
definitely worse than the one in the main panel, is not totally lost
is due to the fact that the behavior $V^{\chi\chi}_{\widehat C}\sim (t-s)^\alpha$
can be written in the scaling form (\ref{scaling}) with $\alpha$ in
place of $a$. Since the exponents $a$ and $\alpha$ are not too
different one would need a much larger range of $s$ in order to
resolve the different curves in the lower inset of
Fig.~\ref{fig_neq_3}.  Notice that for $\widehat C$ small enough,
when the restricted averages scale as in Eq.~(\ref{scaling}) the
scaling function seems to be similar to the one of the whole
$V^{\chi\chi}$, apart from the change in sign. This statement is 
proven by
comparing $-s^{-a}V^{\chi\chi} _{\widehat C}$ with $s^{-a}V^{\chi\chi}$ (dotted
line in the inset).

\subsection{Numerical results and time-reparametrization invariance} \label{numvsrep}

In this Section we re-consider the numerical results 
of Sec.~\ref{EAnumeric} for the spin glass and we discuss whether, how, and to which extent 
they may be interpreted as supporting TRI in the relaxation dynamics of the 
$3d$ EA model. We also report, in Sec.~\ref{comparison} the result of simulations of  
coarsening systems, where TRI is not expected to hold, to signal the differences with the dynamics of 
the disordered model.

\subsubsection{$3d$ Edwards-Anderson model}

A first observation regards the transformation properties 
of the variances $V^{XY}$. 
We have shown in Sec.~\ref{formalism} that if TRI holds
the variances 
must all transform in the same way, analogously to what happens
for the usual two-time functions 
$C$ and $\chi$. According to the discussion of Sec.~\ref{sec:scaling},
when scaling holds Eq.~(\ref{eq:scalVk}) must be obeyed
with a unique exponent $a_{XY}$, independently of the choice of $XY$. 
Our findings, expressed in Eq.~(\ref{scaling}), show that scaling
holds with an algebraic function $\xi (t)$, and that all the variances
scale with the same exponent $a\simeq 0.3$.
This result, therefore,
is in agreement with the predictions of TRI. 
Physically, the scaling (\ref{scaling}) with a single exponent $a _{XY}=a$
amounts to say that, fixing $t/s$, 
the variances of $\widehat C$ and $\widehat \chi$ are 
proportional. This means that basically  
the fluctuations of $\widehat \chi$ are
triggered by those of $\widehat C$ (or vice versa).
As we shall discuss in Sec.~\ref{comparison} the
situation is very different in coarsening ferromagnets.

The behavior of the restricted variances may help us 
better understand  this point.
As expressed above, if TRI holds the non-equilibrium fluctuations of 
$\widehat \chi$ and $\widehat C$
are deeply related, and one can modify the former by manipulating the latter.
Restricted averages are a tool to produce an extreme perturbation of the fluctuations
of $\widehat C$, because we impose $V^{CC}_{\widehat C}=0$.  
As discussed in Sec.~\ref{secneq} this radically changes the behavior
of the fluctuations of $\widehat \chi$, since even the sign of
$V^{\chi\chi} _{\widehat C}$ is the opposite of the one of $V^{\chi\chi}$. 
The change in sign and the fact that 
$V^{\chi\chi}_{\widehat C}$ also scales 
according to Eq.~(\ref{scaling}), imply that
\be
{\cal V}^{\chi\chi} >\sum _{\widehat C}V^{\chi\chi}_{\widehat C} P(\widehat C),
\label{proper_sg}
\ee
from Eq.~(\ref{compos_var}).
This means that the most relevant source of fluctuations of $\widehat \chi$
are the fluctuations of $\widehat C$, while
the fluctuations of $\widehat \chi$ which are independent of the
variations of $\widehat C$ are less important.
Again, this property is not found in coarsening systems, as we shall show
in Sec.~\ref{comparison}.

\subsubsection{Comparison with coarsening systems} \label{comparison}

TRI is not expected to hold in coarsening systems quenched below $T_c$ \cite{ChaCugYos06}
where an infinite effective temperature~\cite{Teff} is developed.
While the features of the $3d$ EA spin-glass discussed insofar 
basically agree with what one would expect in the presence of TRI, 
in this Section we show that the same analysis performed on a phase-ordering
system, the Ising model, leads to a striking disagreement.
  
The behavior of the variances in the Ising model quenched below $T_c$ was 
studied numerically in \cite{CorLipSarZan10f}. 
A scaling form analogous to Eq.~(\ref{scaling}) was found,
with exponents compatible with $a_{CC}=d/2$, $a_{C\chi}=a_{\chi\chi}=1/2$
in $d=1,2$ (and conjectured to hold for higher dimensionality).
The scaling functions $f_{XY}(z)$ decay as $z^{-\lambda_{XY}}$ with 
$\lambda_{CC}=0$, $\lambda_{C\chi}=0$, and 
$\lambda_{\chi\chi}=1/2$.
This shows that, except at the lower critical dimension $d=1$ where the
anomaly of a finite effective temperature is found \cite{LipZan00}, 
the variances scale with different exponents. Moreover, the behavior of the scaling functions
is also very different. As discussed in 
Sec.~\ref{formalism}, this 
excludes the possibility of TRI. 

\begin{figure}
    \centering
   \rotatebox{0}{\resizebox{.7\textwidth}{!}{\includegraphics{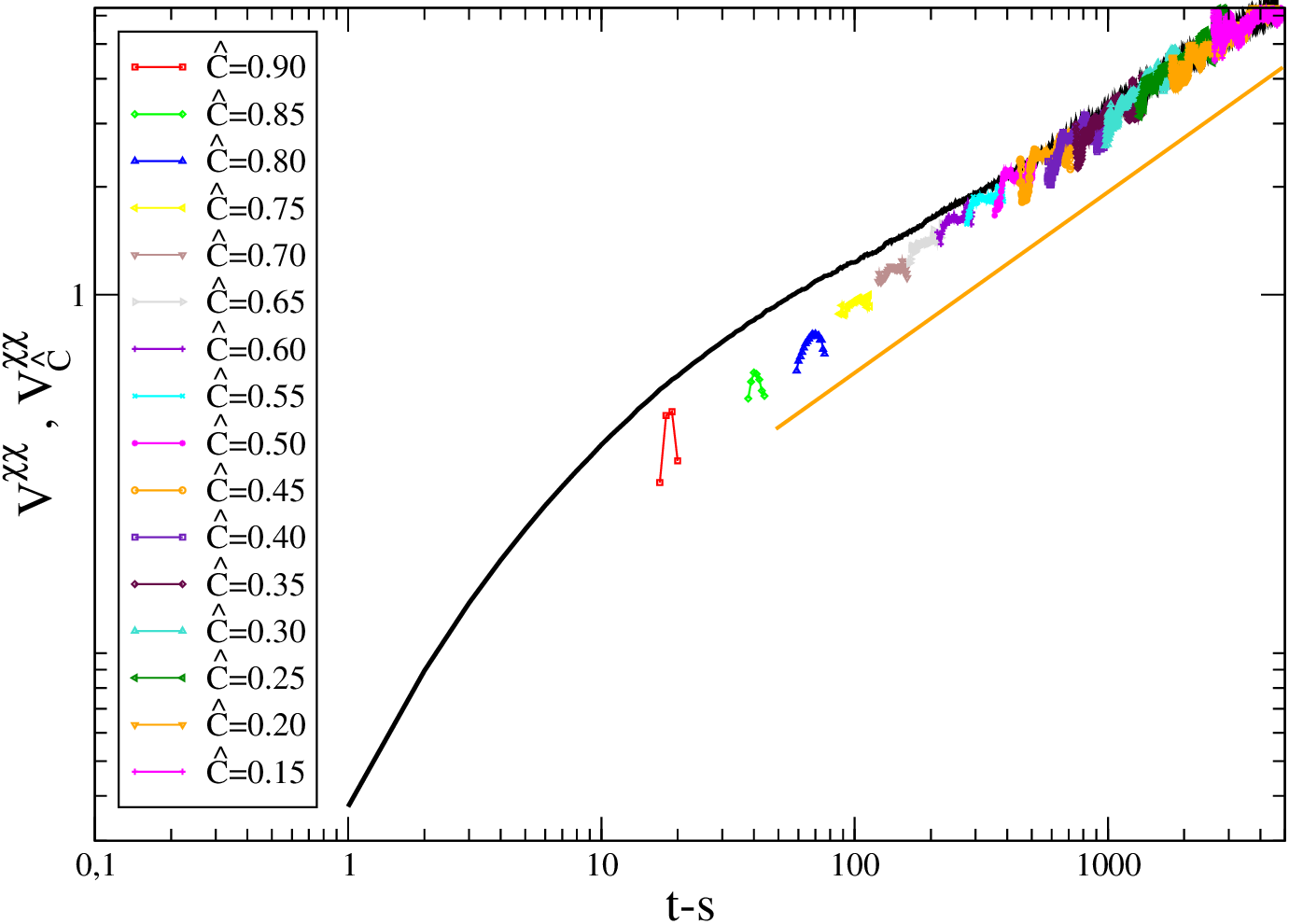}}}
   \vspace{0.5cm}
   \caption{(Color online.) 
   Fluctuating dynamics in the $2d$ ferromagnetic Ising model.
The restricted averages $V^{\chi\chi}_{\widehat C}$ for different values of
$\widehat C$ as a function of $t-s$ with fixed $s=10^2$ MCs 
are shown with (colored) line-points following the code given in the key. 
Similar results are found for other values of $s$. The data are shown in 
a log-log scale. The fully averaged variance $V^{\chi\chi}$ is displayed with a solid black line.
Averages are 
taken over $1.1 \times 10^5$ realizations of the randomness and initial conditions.
The aging regime starts at $\widehat C\simeq 0.65$ where $V^{\chi\chi}_{\widehat C} \simeq
V^{\chi\chi}$.}
\label{fig_neq_ferro}
\end{figure}

The restricted averages were not studied in phase ordering kinetics so far. 
In order to establish how generic the behavior
observed in the $3d$ EA model is, we computed these quantities in coarsening
systems. We simulated the Ising model on a  $d=2$ square lattice, after a quench 
from infinite temperature to $T=0$.
The system size is $L=500$. 
As compared to the spin-glass,
this relatively large value is needed to avoid finite size effects, since
the size of correlated regions grows much faster in this problem.
As a consequence, large fluctuations of $\widehat C$ around the average are
very unlikely. Then, when computing $V^{\chi\chi} _{\widehat C}$ we can collect a significant
statistics only in a narrower time interval than in the spin-glass case.
This can be seen in Fig.~\ref{fig_neq_ferro}, where
the behavior of $V^{\chi\chi}$ and $V^{\chi\chi}_{\widehat C}$ are shown. 
For $V^{\chi\chi}$, we basically reproduce the results in~\cite{CorLipSarZan10f}
(where however a quench to $T=1.5<T_c$ was considered). This quantity 
is positive and behaves as $V^{\chi\chi}\simeq s^{1/2}f_{\chi\chi}(t/s)$,
where $f_{\chi\chi}(z)$ is a function which grows as $z^{1/2}$ for large $z$.
Interestingly, the restricted average $V^{\chi\chi}_{\widehat C}$   
not only has the same sign as the global average, but it superimposes
almost exactly on $V^{\chi\chi}$, for $t-s$ sufficiently large (in the aging regime).
This means that in the non-equilibrium regime $V^{\chi\chi}$ is completely
determined by $V^{\chi\chi}_{\widehat C}$. According to Eq.~(\ref{compos_var}), this implies 
\be
{\cal V}^{\chi\chi} \ll \sum _{\widehat C}V^{\chi\chi}_{\widehat C} P(\widehat C), 
\label{proper_ferro}
\ee
which should be compared to the property (\ref{proper_sg})
found in the $3d$ EA model. 
This indicates that the non-equilibrium fluctuations of $\chi $ 
are entirely produced at fixed $\widehat C$, and are then independent 
of those of $\widehat C$. This behavior, which is qualitatively different 
from
the one observed in the $3d$ EA model, is a clear demonstration of the lack of
TRI in this system.

\vspace{1.5cm}

\section{Summary and conclusions} \label{concl}

This paper is devoted to the study of the properties of fluctuations in aging systems.
Specifically, we concentrate on the fluctuations of the autocorrelation
and dynamic susceptibility, since the relevance of their relations to
the issue of TRI was pointed out 
in~\cite{Chamon02,Castillo02,ChaCugYos06}. 
In the MSRJD path-integral formulation
of Langevin processes, which is overviewed in the first part of the paper,
response functions are associated to the auxiliary field $i\hat \phi$
and hence their fluctuating parts cannot be expressed in terms of
the physical field $\phi $ alone. A possible way out of this limitation
is to consider moments of such fluctuating parts, since these can be related
to generalized response functions which are directly measurable. 
In particular,
in this paper we focused on the {\it variances} $V^{XY}$, because higher
moments are numerically out of reach even with the highly efficient
method adopted here, based on the generalization of the 
fluctuation-dissipation theorem out of equilibrium and beyond
linear order. The behavior of these variances fits in the general
framework of two-time quantities in aging systems. For short time differences
a quasi-equilibrium stationary behavior is observed, while in the
aging regime time-translation invariance is lost and is substituted by a
different, scaling symmetry with a unique exponent $a_{XY}=a$ 
for all the variances. 
In the analytic part of the paper we derived the scaling properties of the
variances and we showed that such a unique exponent is a prediction of TRI. 
The numerical results in the $3d$ EA 
therefore gives support to TRI.  

In order to better characterize  the fluctuations,
we also computed the restricted average $V^{\chi\chi}_{\widehat C}$.
The values of the restriction $\widehat C$ is an additional parameter
that tunes the crossover from the quasi-equilibrium behavior to
the fully aging regime. In particular, the latter is obtained
letting $\widehat C$ be small (with respect to the EA order parameter).
In this case, the restricted average 
scales as the fully averaged one, but its actual value is
not the main contribution to the latter. 
These results have to be compared to the 
ones found in the Ising model. 
In this clean coarsening case the exponents $a_{XY}$ do depend on the 
fluctuating observables and the restricted average $V^{\chi\chi}_{\widehat C}$
is of the same order as the full average $V^{\chi\chi}$,
showing that fluctuations of $\widehat C$ are not the main source
for those of $\widehat \chi$, at variance to what 
it is found in the disordered model. 
These qualitative and quantitative differences between these two systems
add to differences found in the relation between the averaged linear 
response and the self-correlation, i.e. the effective temperature~\cite{Teff}. 

Our results suggest that the fluctuating out of equilibrium dynamics of the 
$3d$ EA model and of coarsening systems are different and that 
this can be related to the fact that TRI is realized in the former model 
while it is not in the latter. This should also be linked to the fact that the effective 
temperature is finite in the former and infinite in the latter cases.

These ideas should be checked numerically and analytically in other 
out of equilibrium systems. Natural cases to analyze are the 
dynamics of atomic glass models (like soft sphere, Lennard-Jones 
mixtures or other) with molecular dynamics or the evolution of kinetically 
constrained models~\cite{kinetically-constrained}. The Glauber Ising chain is an example that 
could be dealt with, possibly, analytically, exploiting the techniques 
derived in~\cite{Mayer}.
Studies of critical 
dynamics~\cite{AnnSol09} and of the relaxation of
a model at its lower critical dimension should help to complete this picture.

\vspace{1cm}

We thank E. Lippiello, A. Sarracino, and M. Zannetti for many useful 
discussions. 
F.~C. acknowledges financial support from PRIN 2007JHLPEZ 
({\it Statistical Physics of Strongly correlated systems in Equilibrium
and out of Equilibrium: Exact Results and Field Theory methods}) and from 
Universit\'e Pierre et Marie Curie and thanks the LPTHE Jussieu for 
hospitality during the preparation of this work. C.~C. and L.~F.~C. thank the Universidad de 
Buenos Aires, Argentina, for hospitality during the last stages of this work.
This work was supported in part by DOE Grant DEFG02-06ER46316 (C.~C.).

\end{document}